\documentclass[conference]{IEEEtran}
\usepackage{url}
\usepackage{color}
\usepackage{graphicx}
\usepackage{epstopdf}
\usepackage{amsmath}
\usepackage{gensymb}
\usepackage{amssymb}
\usepackage{mathtools}
\usepackage{array}
\usepackage{booktabs}
\usepackage{multirow}
\usepackage{threeparttable} 
\usepackage{comment}
\usepackage{soul}
\usepackage{hyperref}
\PassOptionsToPackage{bookmarks=false}{hyperref}
\hypersetup{draft}

\usepackage{graphicx}
\usepackage{xcolor}
\usepackage{listings}

\lstdefinestyle{BashInputStyle}{
  language=bash,
  basicstyle=\scriptsize\sffamily,
  numbers=none,
  numberstyle=\tiny,
  numbersep=3pt,
  frame=tb,
  columns=fullflexible,
  backgroundcolor=\color{yellow!20},
  linewidth=\linewidth,
  xleftmargin=0\linewidth
}

\IEEEaftertitletext{\vspace{-3\baselineskip}}

\usepackage{setspace}

\usepackage{enumitem}
\setlist{nosep} 

\newcommand\blfootnote[1]{%
  \begingroup
  \renewcommand\thefootnote{}\footnote{#1}%
  \addtocounter{footnote}{-1}%
  \endgroup
}

\begin{document}
\title{Leaving Your \textit{Things} Unattended is No Joke! {\huge Memory Bus Snooping and Open Debug Interface Exploits}}

\author{
   \IEEEauthorblockN{Yang Su and Damith C.~Ranasinghe}
  \IEEEauthorblockA{\textit{Auto-ID lab, School of Computer Science, The University of Adelaide}, Australia
  \\ \{yang.su01; damith.ranasinghe\}@adelaide.edu.au}
}

\maketitle
\thispagestyle{plain}
\pagestyle{plain}
\blfootnote{© 2022 IEEE.  Personal use of this material is permitted.  Permission from IEEE must be obtained for all other uses, in any current or future media, including reprinting/republishing this material for advertising or promotional purposes, creating new collective works, for resale or redistribution to servers or lists, or reuse of any copyrighted component of this work in other works.}
\begin{abstract}
Internet of Things devices are widely adopted by the general population. People today are more connected than ever before. The widespread use and low-cost driven construction of these devices in a competitive marketplace render Internet-connected devices an easier and attractive target for malicious actors. This paper demonstrates non-invasive physical attacks against IoT devices in two case studies in a tutorial style format. The study focuses on demonstrating the: i)~exploitation of debug interfaces, often left open after manufacture; and ii)~the exploitation of exposed memory buses. We illustrate a person could commit such attacks with entry-level knowledge, inexpensive equipment, and limited time (in 8 to 25 minutes).
\end{abstract}


\section{Introduction}
A considerable amount of IoT devices have been adopted by consumers and industries over the last decades; the strong growth momentum is set to continue. In 2020, the number of Internet-connected smart devices reached 11.7 billion and is expected to increase by more than five-fold by 2025~\cite{afuang_rago_mukherjee_rojas_ujhazy_2020}. Apart from the popular consumer electronics, health care, and manufacturing, the automotive industry is the next promising market, followed by retail, logistics, agriculture, and animal husbandry sectors~\cite{liyanage2020iot}. Despite the rapid growth, there are still many challenges related to security, especially for devices involved personal and sensitive data. By 2025, the amount of data gathered by those devices are expected to reach 73.1~zettabytes~\cite{afuang_rago_mukherjee_rojas_ujhazy_2020}. The widespread existence and access to sensitive data render Internet-connected devices a lucrative attack target for malicious actors~\cite{khelif2021non}. 

Despite user privacy concerns, it is difficult to choose products with adequate protections without an in-depth knowledge on security and privacy~\cite{badran2019iot}. Meanwhile, manufacturers have generally focused on reducing time-to-market delays~\cite{jha2014security}, improving  cost-effectiveness and quality of service~\cite{hammi2020secure} but skipping over some necessary security measures~\cite{capellupo2017security}.

\vspace{1mm}
\noindent\textbf{Our Focus and Contributions.~}This paper will focus on investigating the vulnerabilities arising from common debug interfaces left open and exposed memory buses in IoT (Internet of Things) devices. Notably, unlike software that can be patched, we focus on exploiting vulnerabilities related to the hardware design that are difficult to be fixed after products are released to the market. To demonstrate the ease with which snooping\footnote{\scriptsize Snooping: unauthorized access, similar to eavesdropping but not limited to data collection during transmissions.} on exposed memory buses and exploiting open debug interfaces can be, we apply techniques that are simple in practice, only require inexpensive gadgets, and do not impose irrecoverable or noticeable damage to the device. In particular, in this article, we: 
\begin{itemize}
    \item Investigate the practical threat posed by: i)~debug interfaces left open; and ii)~exposed buses to off-chip memory commonly used for storing data and secrets in COTS (commercial off-the-shelf) devices.
    \item Demonstrate the relatively low cost and the level of skills required to extract sensitive data from devices. 
    \item Conduct and describe two case studies, in \textit{the style of a tutorial}, to demonstrate the realistic threat by extracting executable code and secrets from COTS electronic devices. A demonstration video of gaining remote access to a portable Internet connected camera after an exploit is available at \underline{https://youtu.be/fnIn9QugrXI}.
\end{itemize}

\vspace{1mm}
\noindent\textbf{Organisation.~}The rest of the paper is organized as follows. In Section~\ref{sec:RelatedWk_n_resources}, we summarize previous work and useful resources related to physical attacks on IoT devices. Section~\ref{sec:threat_model} presents the threat model. Exploitable debug interfaces and memory buses are discussed in Section~\ref{sec:exploitable_buses} followed by two case studies using popular IoT devices. Section~\ref{sec:conclusion} provides concluding remarks.

\section{Background} \label{sec:RelatedWk_n_resources}
\noindent \textbf{Software attacks.~} Attacks against software exploit security vulnerabilities of the communication protocols, cryptographic algorithms, or software implementation of the product~\cite{skorobogatov2012physical,davis2020vulnerability,sasaki2009finding}. Software attacks are possible without physical access to the victim device. However, the manufacturer could easily fix the software vulnerabilities through system updates~\cite{morel2019idols,su2021wisecr}. In this paper, we would like to focus on exploiting vulnerabilities related to the hardware design that are difficult to be fixed after products are released to the market.

\vspace{1mm}
\noindent \textbf{Invasive physical attacks.~} These attacks require direct access, for example, dissect a smart card and the use of microprobes to read the secure storage of cryptographic keys~\cite{yuan2019fast} from its internal components~\cite{skorobogatov2012physical,ibrahim2018us}. We consider methods that obviate invasive attacks since it generally requires expensive equipment (e.g., a US\$5,000+ microprobe workstation) and, inevitably, leaves traces of tampering with the device~\cite{pammu2016interceptive}. 

\vspace{1mm}
\noindent \textbf{Non-invasive physical attacks.~} Vasile et al.~\cite{vasile2018breaking} summarized three major firmware extraction techniques: i)~Debug interfaces; ii) Raw Flash Dump; and Software Methods. The author also examined 24 popular COTS smart devices in 2018, and 100\% of them are vulnerable to at least one of the techniques, followed by three case studies. This work also proposed countermeasures for each of the exploits. Although the study is inspiring and covers a wide spread of target devices, the focus of the study was the techniques to dump the firmware from the devices---leading to intellectual property theft. However, exploitation or what useful information the dumped firmware could provide, time and monetary cost of an attack or a practical adversary model extracting the firmware and the consequence of having access to that firmware was unclear. Krishnan and Schaumont, in~\cite{krishnan2018exploiting} investigated exploiting the JTAG (Joint Test Action Group) interface in an intermittent computing system. They demonstrated the extraction of the AES (Advanced Encryption Standard) secret key from checkpoints stored in on-chip NVM (non-volatile memory). However, in their adversary model, an active attacker has unrestricted physical access and can corrupt and modify the target device's memory content. 

\section{Threat Model}\label{sec:threat_model}
We build our threat model on the basic assumption that without technical knowledge, ordinary users would not treat IoT devices without obvious mark of tampering as malicious and warrant further investigation. The attacker's goal is to extract sensitive information from the IoT device hardware, including but not restricted to the user's personal information, usage data, device configuration and executable code. 

\subsection{Victim Devices}
We classify the victim IoT devices into: i)~\textbf{Class-I}: The device implements debugging or programming interfaces, and such interfaces were left open in the final product; and ii)~\textbf{Class-II}: The devices store sensitive data in off-chip NVM, and the memory buses are exposed at the top/bottom layer of the PCB (Printed circuit board) in the final product.


\subsection{Attacker Capabilities}
\begin{itemize}
    \item Has physical procession to the victim device for a restricted time window---a few minutes to half an hour---such a short time generally does not attract the device owner's attention. Notably, this setting renders it impractical for physical attacks that require a longer time~\cite{ibrahim2018us} for specialized laboratory setups and data collection.
    \item Has a basic understanding of electronics, computer security, knows how do debug interfaces and serial buses function. It is important to highlight that it is easy to acquire the simple knowledge needed from materials freely available online, such as~\cite{bartlett2020electronics} and~\cite{gupta2019iot}.
    \item Has access to inexpensive tools as exemplified in Section~\ref{sec:attacker_tool} and a computer with open-source decryption and binary file analysis programs.
\end{itemize}
Our attacker capabilities are reasoned based on~\cite{krishnan2018exploiting}, with following changes to consider more generic threats: 1) the duration of attacker's physical access is restricted; 2)~an attack does not leave visible traces (e.g., modify the configuration or data) to avoid drawing a user's attention; 3)~we obviated attacks that requires expert knowledge or expensive equipment.

\subsection{Resources and Costs}\label{sec:attacker_tool}
We assume an attacker has access to the tools listed in Appendix~\ref{apd:attacker_tools} to facilitate memory bus snooping and open debug interface exploits. The reference prices were obtained from \underline{www.ebay.com}. For the attacks we explore, the total cost of a memory bus exploit attack is US\$15, and the cost of an open JTAG interface exploit attack is US\$45. We can observe that an attack under this threat model is inexpensive, and tools can be easily acquired to facilitate memory bus snooping and open debug interface exploits. Notably, an attack under the settings does \textit{not} leave irreversible or visible damage to the device~\cite{pammu2016interceptive} or attempt to modify firmware (as that could permanently brick the device~\cite{cui2013firmware,morel2019idols}).

\section{Exploitable Debug Interfaces and Exposed Memory Buses}\label{sec:exploitable_buses}
This section will introduce exploitable debug interfaces and memory buses in COTS IoT devices and demonstrate how we can easily identify these interfaces for exploitation.

\subsection{Open Debug Interfaces}
\begin{figure}[!ht]
    \centering
    \includegraphics[width=\linewidth]{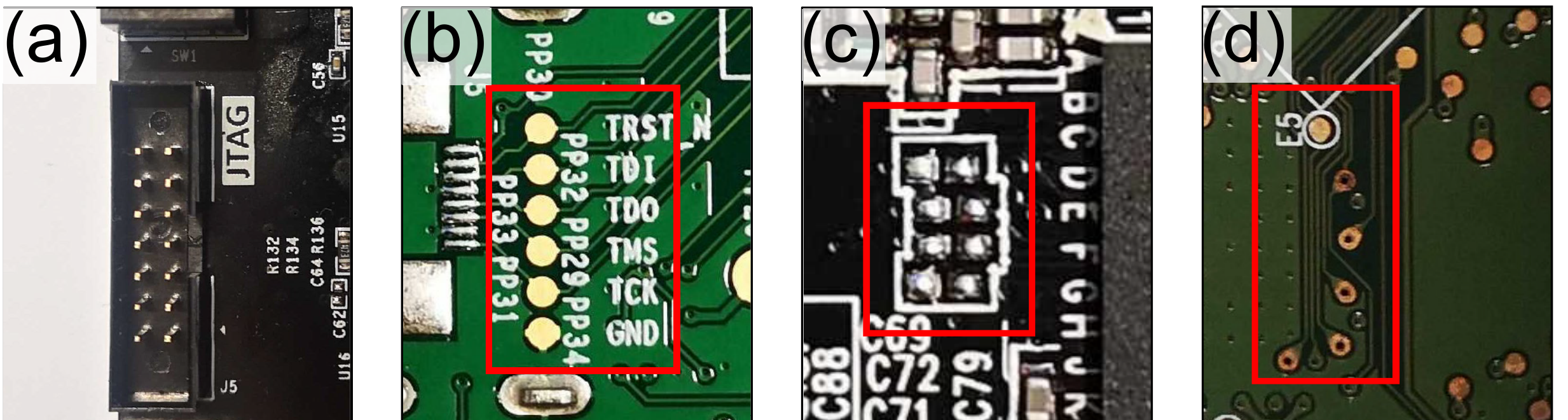}
     \caption{JTAG debug interface found in IoT devices: (a) labeled with JTAG silk print; (b) labeled with JTAG pin-outs (e.g., \texttt{TDI}, \texttt{TDO}, \texttt{TMS} or \texttt{TCK}); (c) one or two row of test points nearby the MCU; and (d) hidden JTAG.}
    \label{fig:JTAG_interfaces}
\end{figure}

JTAG is a commonly used debug interface for embedded systems. We may find one or two rows of jumper pins with a printed JTAG, JT\textit{x} or J\textit{x} label, where \textit{x} might be a number as in \autoref{fig:JTAG_interfaces}~(a). Occasionally, we can also find individual pins labelled with \texttt{TDI}, \texttt{TDO}, \texttt{TMS} or \texttt{TCK} as seen in \autoref{fig:JTAG_interfaces}~(b). In some products, JTAG headers are there without any label--- \autoref{fig:JTAG_interfaces}~(c)---or even hidden from sight as seen in~\autoref{fig:JTAG_interfaces}~(d).

\begin{table}[!ht]
\caption{Basic JTAG signals. A JTAG interface may also have \texttt{Vsupply}, \texttt{VTref}, \texttt{RTCK} and \texttt{TRACE} signals---as described in the user manual of a JTAG debugger.}
\label{tab:JTAG_signals}
\centering
\resizebox{.9\linewidth}{!}{%
\begin{tabular}{lll}
\toprule
Signal name & Direction & Description \\
\midrule
\texttt{TRST\_N} &  Programmer $\rightarrow$ Device & JTAG reset (active low) \\ 
\texttt{TDI} &  Programmer $\rightarrow$ Device & JTAG scan-chain input \\ 
\texttt{TDO} &  Programmer $\leftarrow$ Device & JTAG scan-chain output \\ 
\texttt{TMS} &  Programmer $\rightarrow$ Device & JTAG mode selection \\ 
\texttt{TCK} &  Programmer $\rightarrow$ Device & JTAG clock \\ 
\texttt{GND} &  Programmer $\leftrightarrow$ Device & Common ground\\ \bottomrule
\end{tabular}%
}
\end{table}

To use the basic functionality of the JTAG debug interface, the five signals shown in \autoref{tab:JTAG_signals} need to be wired up correctly to a JTAG Debugger---see \autoref{fig:JTAG_lock_attched}. There are no concrete definitions of the pin-out order for the JTAG header. A manufacturer can place the JTAG pins in any order. To determine the JTAG pin-definition, we can use the manual inspection method in \autoref{sec:jtag_case_study} or use the open-source JTAGenum described in Appendix~\ref{apd:attacker_tools}.

\begin{figure}[!ht]
    \centering
    \includegraphics[width=\linewidth]{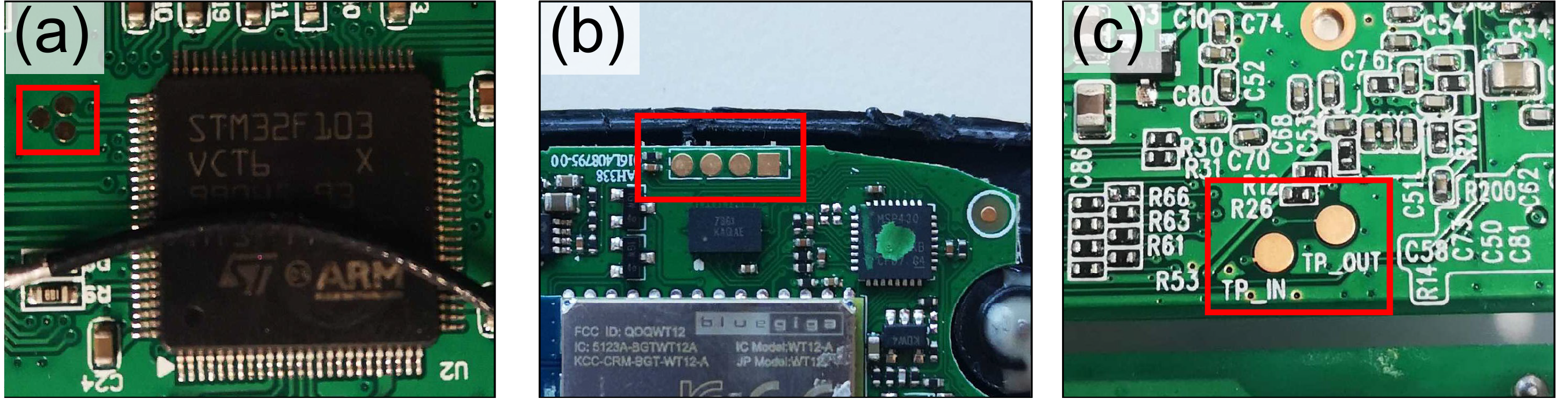}
     \caption{Two-wire debug interfaces found in IoT devices: (a) ARM SWD debug interface; (b) TI SBW debug interface; and (c) manufacturer defined proprietary two-wire debug interface.}
    \label{fig:SWD_interfaces}
\end{figure}

Other than JTAG, manufacturers may implement different debug interfaces, such ARM SWD~(Serial Wire Debug in~\autoref{fig:SWD_interfaces}~(a)) and TI SBW~(Spy-Bi-Wire in~\autoref{fig:SWD_interfaces}~(b)). IoT device manufactures can also implement a proprietary two-wire debug interface (e.g., the that found in a TP-Link WiFi access point~\autoref{fig:SWD_interfaces}~(c). The two-wire programming interfaces typically consist of one bi-directional data wire and one clock wire, sometimes including GND~(ground) and Vcc~(common collector voltage) for potential reference and power supply. Two-wire debug interfaces benefit from the reduced pin design and are suitable for IoT devices with small form factors.

\subsection{Exposed Memory Buses}
Since the on-chip NVM is limited and expensive~\cite{li2019chip}, many IoT devices use off-chip Flash or EEPROM (electrically erasable programmable read-only memory) to store firmware images, settings, and secrets.

\begin{figure}[!ht]
    \centering
    \includegraphics[width=\linewidth]{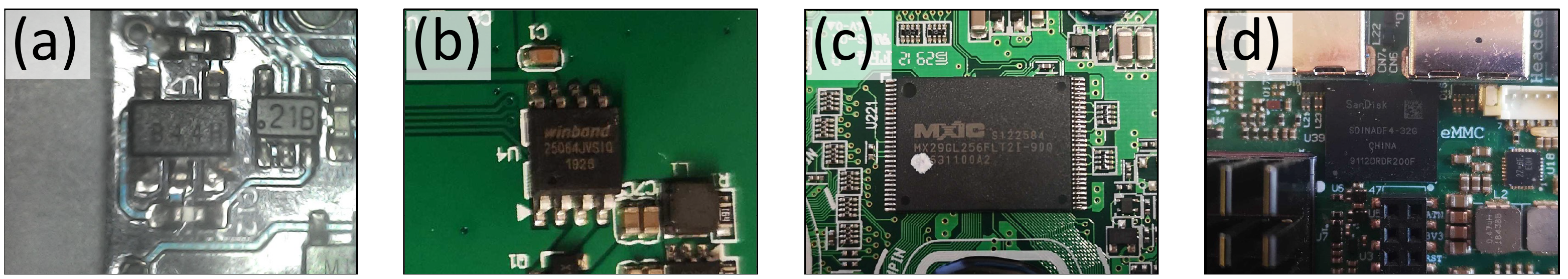}
     \caption{Different NVM chips in IoT devices: (a) EEPROM in SOT23-5 package; (b) Flash in SOP-8 package; (c) Flash in TSOP-56 package and (d) Flash in eMMC-153b package.}
    \label{fig:NVM_chip}
\end{figure}

\autoref{fig:NVM_chip} lists a number of off-chip memory in IoT devices. EEPROM in SOT12-5 package in \autoref{fig:NVM_chip}~(a) often appears in tiny devices, such as a smartwatch, due to its small footprint. Flash or EEPROM chips in the SOP-8 package are common off-chip storage in many IoT devices; such memory communicates with the microcontroller via a serial bus such as I2C~(Inter-Integrated Circuit) and SPI~(Serial Peripheral Interface), as exemplified in \autoref{fig:NVM_chip}~(b). IoT devices requiring relatively higher storage space and faster access speeds may employ Flash chips with parallel buses, such as those shown in \autoref{fig:NVM_chip}~(c) and (d)---parallel buses employ more wires to transfer data and address signals.  We focus on memory chips using serial buses, as parallel buses are generally difficult to access and require professional and costly equipment. In addition, memory chips with parallel buses can use a BGA (ball grid array) package where all signal pins are hidden behind the package; therefore, it is more difficult to access~\cite{vasile2018breaking}. 

\section{Open JTAG Interface Exploit Case study}\label{sec:jtag_case_study}
Post identification of interfaces for possible access, this section investigates the potential threat posed by demonstrating the extraction of memory contents in IoT devices in two example case studies under the practical threat model we consider in Section~\ref{sec:threat_model}.

\vspace{1mm}
\noindent\textbf{Scenario.~}Consider the following scenario for our first case study: \textit{Elizabeth is an 87-year-old widow, she lives alone in an apartment equipped with an electronic lock. One day the electronic lock runs out of battery. Elizabeth asked one of her neighbors to replace the battery for the lock. However, the battery in the lock is an unusual model. The neighbor suggests bringing the lock to the hardware store to match the correct battery model. Half an hour later, the neighbor returned with the battery and re-installs the electronic lock for Elizabeth.} Could the neighbor extract the key code for the electronic lock within the 30-minute time window to gain unauthorized access to Elizabeth's house?

\begin{figure*}[!ht]
    \centering
    \includegraphics[width=\linewidth]{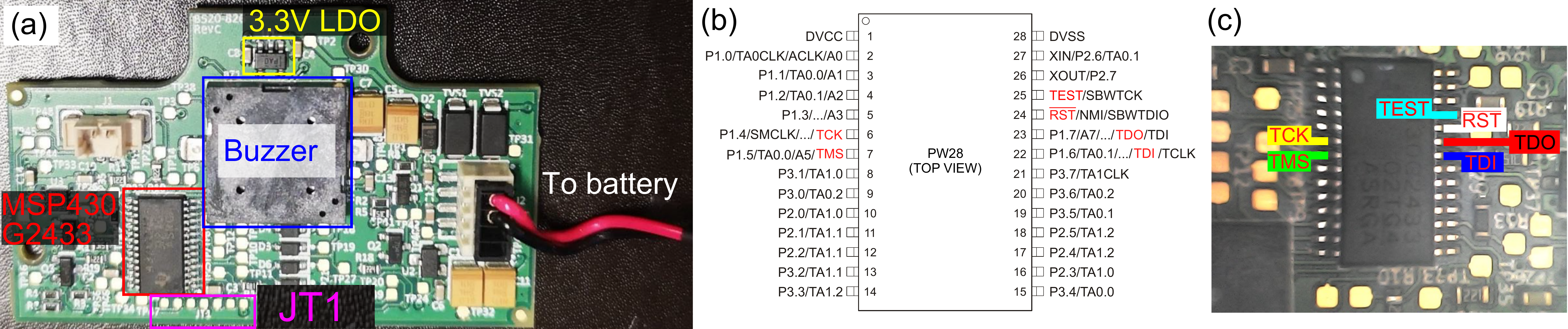}
     \caption{The main board (a) of the electronic lock. The embedded microcontroller in the lock is a MSP430G2433, and JT1 is highly likely to be the JTAG interface. Highlighted JTAG signals on (b) are pin definitions reproduced from MSP430G2433 datasheet~\cite{instrumentsmsp430G2x33}; and (c) the microcontroller on the main board.}
    \label{fig:Lock_cuicuit_board}
\end{figure*}

\vspace{1mm}
\noindent\textbf{Attack.~}We use a Schlage electronic lock FE575 as an example. The tear-down of the electronic lock is summarized in Appendix~\autoref{fig:Lock_teardown}. The mainboard of the Schlage electronic lock FE575 is shown in~\autoref{fig:Lock_cuicuit_board}. Its circuit layout is uncomplicated. The black square in the middle is a buzzer. It makes a sound when the keyboard is pressed. At the left bottom corner of the buzzer is an MSP430G2433 microcontroller-- the electronic lock's brain. Other components are mostly responsible for powering, motor operation---the circuit needs to drive a DC motor to release the lock when a correct password is pressed. More importantly, there are seven pins in a row labeled with JT1, which is highly like a JTAG interface.

Unfortunately, the pin definitions for JT1 are not labelled on the PCB; hence a strategy is needed to identify the correct order of JTAG signals. Fortunately, as seen in~\autoref{fig:Lock_cuicuit_board}, the MSP430G2433 microcontroller is in an SOP-28 package, all pins are easily accessible. Hence a US\$10 multimeter is adequate to identify pin definitions of JT1 without using professional tools such as the US\$200 JTAGulator.

Our technique to find correct JTAG pins is to use the diode and continuity mode of a multimeter. The diode and continuity mode is commonly labeled with a diode and a sound wave symbol (e.g., \includegraphics[height=1.2\fontcharht\font`\B]{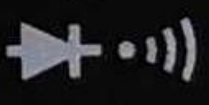} ). This mode measures the forward bias of a diode (if there is one between the two probes of the multimeter). If there is a direct wire connection (or short circuit), the multimeter buzzer will sound. Depending on the multimeter manufacturer, different probe polarities in diode test mode may be used. For our model Stanley STHT77364, the red probe is connected to the cathode, and the black probe is connected to the anode of the tested diode (although, typically, red is the anode and black is the cathode, refer to the multimeter model used for probe polarities). We will start with seeking for the \texttt{GND}, which can be easily accessed at: i)~The negative terminal of the battery; ii) any EMI (Electromagnetic interference) shielding used; iii)~any pin connected to the ground copper pour; iv)~metal case of connectors, such as USB; v)~the \texttt{GND} or \texttt{Vss} pin of a known IC.

Plug the probes into the correct receptacle of the multimeter, select the diode and continuity mode. Attach the red multimeter probe to any of the exposed \texttt{GND} listed above. We choose to use the negative terminal of the buzzer. As this terminal is relatively large, it is easy to attach the probe and is directly connected to the ground copper pour. Then use the black multimeter probe to explore each of the seven pins in JT1. As summarized in~\autoref{tab:JT1_table} under the column \texttt{GND}, the $5^{th}$ pin of the JT1 has forward-biased voltage drop $V_F = 0$, which implies the JT1.5 is the \texttt{GND} pin in this JTAG connector.

Other JTAG signals can be determined using a similar method. By looking up the datasheet of the MSP430G2433 microcontroller, we have highlighted the six JTAG signal pins  in~\autoref{fig:Lock_cuicuit_board}. Now attach the multimeter's black probe to each of the five highlighted microcontroller leads, and measure the $V_F$ between them and each JT1 pin. The results are summarized in~\autoref{tab:JT1_table}.

\begin{table}[!ht]
\caption{The forward-biased voltage drop (in mV) measured between each JTAG signal and JT1 pins. Here, OL stands for open loop.}
\label{tab:JT1_table}
\centering
\resizebox{.9\linewidth}{!}{%
\begin{tabular}{cccccccc}
\toprule
\multicolumn{1}{c}{} & \multicolumn{1}{c}{\texttt{GND}} & \multicolumn{1}{c}{\texttt{TRST\_N}} & \multicolumn{1}{c}{\texttt{TDI}} & \multicolumn{1}{c}{\texttt{TDO}} & \multicolumn{1}{c}{\texttt{TMS}} & \multicolumn{1}{c}{\texttt{TCK}} & \multicolumn{1}{c}{\texttt{TEST}} \\ \midrule
\multicolumn{1}{c|}{JT1.1} & \multicolumn{1}{c}{687} & \multicolumn{1}{c}{OL} & \multicolumn{1}{c}{OL} & \multicolumn{1}{c}{OL} & \multicolumn{1}{c}{OL} & \multicolumn{1}{c}{\textbf{0}} & \multicolumn{1}{c}{OL} \\ 
\multicolumn{1}{c|}{JT1.2} & \multicolumn{1}{c}{714} & \multicolumn{1}{c}{OL} & \multicolumn{1}{c}{OL} & \multicolumn{1}{c}{OL} & \multicolumn{1}{c}{\textbf{0}} & \multicolumn{1}{c}{OL} & \multicolumn{1}{c}{OL} \\ 
\multicolumn{1}{c|}{JT1.3} & \multicolumn{1}{c}{711} & \multicolumn{1}{c}{OL} & \multicolumn{1}{c}{\textbf{0}} & \multicolumn{1}{c}{OL} & \multicolumn{1}{c}{OL} & \multicolumn{1}{c}{OL} & \multicolumn{1}{c}{OL} \\ 
\multicolumn{1}{c|}{JT1.4} & \multicolumn{1}{c}{714} & \multicolumn{1}{c}{OL} & \multicolumn{1}{c}{OL} & \multicolumn{1}{c}{\textbf{0}} & \multicolumn{1}{c}{OL} & \multicolumn{1}{c}{OL} & \multicolumn{1}{c}{OL} \\ 
\multicolumn{1}{c|}{JT1.5} & \multicolumn{1}{c}{\textbf{0}} & \multicolumn{1}{c}{715} & \multicolumn{1}{c}{OL} & \multicolumn{1}{c}{OL} & \multicolumn{1}{c}{478} & \multicolumn{1}{c}{450} & \multicolumn{1}{c}{430} \\ 
\multicolumn{1}{c|}{JT1.6} & \multicolumn{1}{c}{716} & \multicolumn{1}{c}{\textbf{0}} & \multicolumn{1}{c}{OL} & \multicolumn{1}{c}{OL} & \multicolumn{1}{c}{OL} & \multicolumn{1}{c}{OL} & \multicolumn{1}{c}{OL} \\ 
\multicolumn{1}{c|}{JT1.7} & \multicolumn{1}{c}{523} & \multicolumn{1}{c}{OL} & \multicolumn{1}{c}{OL} & \multicolumn{1}{c}{OL} & \multicolumn{1}{c}{OL} & \multicolumn{1}{c}{OL} & \multicolumn{1}{c}{\textbf{0}} \\ \bottomrule
\end{tabular}%
}
\end{table}

\begin{table}[!ht]
\caption{The estimated JT1 pins definitions.}
\label{tab:JT1_estimated}
\centering
\resizebox{.9\linewidth}{!}{%
\begin{tabular}{ccccccc}
\hline
 \multicolumn{1}{c}{JT1.1} & \multicolumn{1}{c}{JT1.2} & \multicolumn{1}{c}{JT1.3} & \multicolumn{1}{c}{JT1.4} & \multicolumn{1}{c}{JT1.5} & \multicolumn{1}{c}{JT1.6} & \multicolumn{1}{c}{JT1.7} \\ \hline
 \multicolumn{1}{c}{\texttt{TCK}} & \multicolumn{1}{c}{\texttt{TMS}} & \multicolumn{1}{c}{\texttt{TDI}} & \multicolumn{1}{c}{\texttt{TDO}} & \multicolumn{1}{c}{\texttt{GND}} & \multicolumn{1}{c}{\texttt{TRST\_N}} & \multicolumn{1}{c}{\texttt{TEST}} \\ \hline
\end{tabular}%
}
\end{table}

Based on the results in~\autoref{tab:JT1_table}, we can conclude that JT1.1 is the \texttt{TCK} signal, as there is a direct wire connection between the two points. Similarly, JT1.2 is \texttt{TMS}; JT1.3 is \texttt{TDI}; JT1.4 is \texttt{TDO}; JT1.5 is \texttt{GND} and JT1.6 is \texttt{TRST\_N}. The JT1.7 is directly connected to the $25^{th}$ pin of the MSP430G2433. According to the datasheet, this is the \texttt{TEST} pin connected to the internal device protection fuse. The product manufacturer should blow the internal fuse by applying a 6~V, 100~mA current to prevent further JTAG access. In addition, \texttt{Vcc} is absent in JT1; in this case, the board is powered from a battery while debugging instead of the JTAG debugger. The estimated JT1 pin definitions are summarized in~\autoref{tab:JT1_estimated}.

\begin{figure}[!ht]
    \centering
    \includegraphics[width=\linewidth]{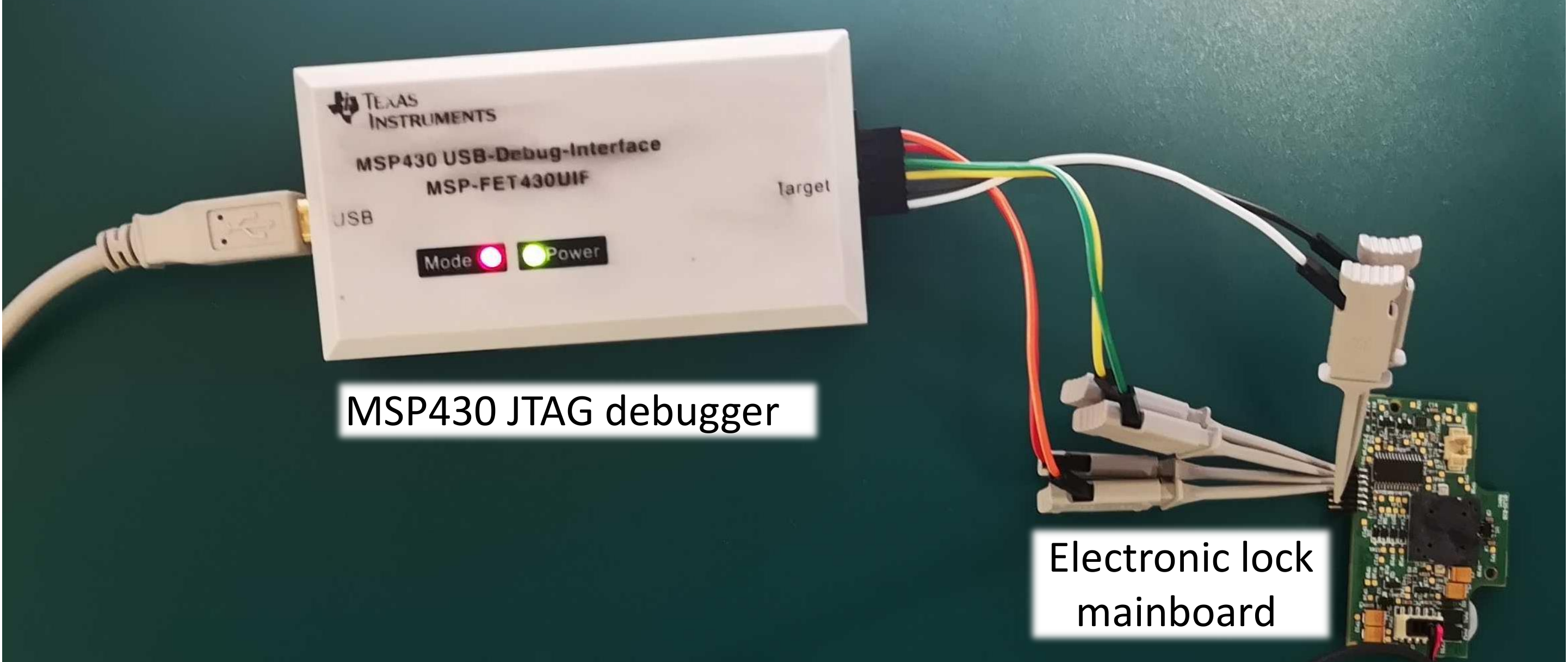}
     \caption{JTAG debugger attached to the electronic lock main board.}
    \label{fig:JTAG_lock_attched}
\end{figure}

Next, we use a TI MSP430 JTAG debugger (or a lower cost, compatible model from a third party) and connect to the pins identified above. We opted to solder a connector to the PCB and use jumper wires to connect to the JTAG debugger as shown in~\autoref{fig:JTAG_lock_attched}. Soldering is not compulsory, a data repair tool (typically costing US\$20) can be used instead.

\begin{figure}[!ht]
    \centering
    \includegraphics[width=\linewidth]{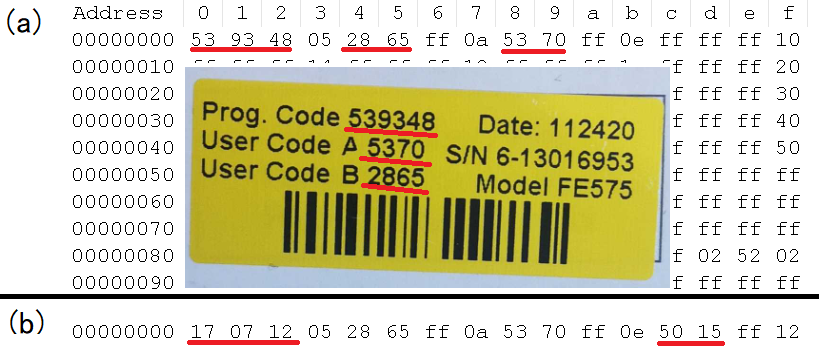}
     \caption{The memory content read from the electronic lock, (a) contains secret programming code (``539348'') and user codes (``5370'' and ``2865''), in plain-text. (b) after the user change the programming code to ``170712'' and added a new user code ``5015'', the new codes can still be read out.}
    \label{fig:Lock_user_code}
\end{figure}
To access the internal memory contents, we used Texas Instrument's UniFlash software available at: \\
\underline{https://www.ti.com/tool/download/UNIFLASH}. If a suitable JTAG debugger is selected to match the target system, the connections are correct, and the JTAG fuse was not blown by the manufacturer, UniFlash will automatically detect the chip model. As shown in~\autoref{fig:Lock_user_code}~(a) the memory content read out from information memory (0x1000-0x10FF according to~\cite{instrumentsmsp430G2x33}), contains the lock's programming code, and user codes. All stored in plaintext. Those codes match that printed on the user's manual of the electronic lock and should be kept secret. Even if the user has changed the default codes, an adversary can still extract the new value from the same address. With the user code, one can gain entry without letting the homeowner know. This raises important questions regarding the security of such devices to insider attacks, and in a more practical setting, leaving the house unattended and in the company of individuals with lower degrees of trust---e.g. short term rental settings. Simple mitigation would be to change the default programming code and blow the fuse by putting a 6~V by 100~mA current to the \texttt{TEST} pin of JT1 to prevent further JTAG access.

\section{Memory Bus Snooping Case study}\label{sec:mem_bus_case_study}
In the second case study, we consider the following.

\vspace{1mm}
\noindent\textbf{Scenario.~}\textit{Emiko is an international student who lives in a shared house with a few housemates. She bought a WiFi IP camera to monitor her room when she was out. One day, her suburb experienced power outages, and Emiko decided to stay at the University until power was restored.} We demonstrate the risk of a malicious actor, under our treat model, sneaking into Emiko's room (when the camera will not raise any potential alarms, given the power outage) and access the WiFi IP camera to gain control and potentially facilitate peeping whilst leaving no visible marks of tampering and a fully operational device.

\vspace{1mm}
\noindent\textbf{Attack.~}We employed a TP-link Tapo C100 IP camera as shown in~\autoref{fig:TapoC100}~(a) to illustrate an attack that exploits an exposed memory bus. The casing of the camera is held together by snap-fit joints; we could disassemble it with a simple lever without leaving irrecoverable damage as shown in~\autoref{fig:TapoC100}~(b). Most of the important logic components are located at the backside of the mainboard, as shown in~\autoref{fig:TapoC100}~(c). The brain of the IP camera is the Realtek SoC (System-on-chip) RTS3903. Unfortunately, its datasheet is not publicly available. Besides the RTS3903, there is an 8~MiB SPI Flash chip XM25QH64C. In this case study, we target extracting information stored in this Flash chip. The XM25QH64C Flash chip is in an SOP-8 package. All its pins are exposed, pin definitions, reproduced from~\cite{wuhan_2020}, are shown in~\autoref{fig:TapoC100}~(d). As we demonstrate, we can easily use low-cost tools to access the exposed memory bus.

\begin{figure}[!ht]
    \centering
    \includegraphics[width=\linewidth]{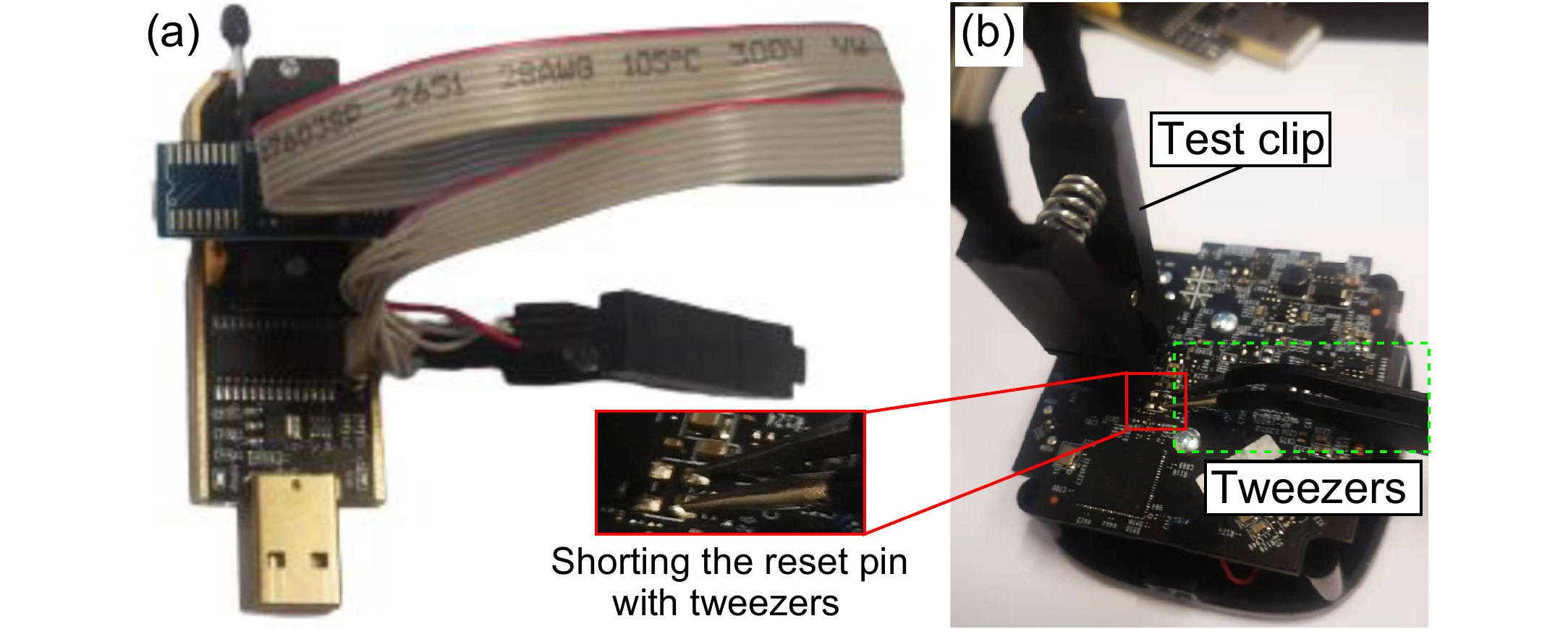}
     \caption{(a) the CH341A USB Flash memory programmer; (b) the test clip of the memory programmer attached to the the MX25QH64 SPI Flash chip.}
    \label{fig:Camera_clip}
\end{figure}

To snoop on the Flash memory chip, the easiest way is to use a US\$10 CH341A programmer with a test clip as shown in~\autoref{fig:Camera_clip}. The programmer will power the Flash chip and override the SPI bus to send access commands, even when the camera is powered off. The memory contents can be dumped out using freely available software\footnote{\scriptsize Available:~\underline{https://www.instructables.com/CH341A-Programmer/}}. 

The IP camera we studied has the Flash chip and the Realtek SoC powered from the same power rail. When the test clip powers the Flash chip, the Realtek SoC will also start up and attempt to access the Flash chip. This will interfere with our memory readout. The easiest way to prevent the interference is to keep the Realtek SoC in a reset state. We spotted four unpopulated connectors near the SoC by inspecting the IP camera circuit board. Using the multimeter, we can conclude pin~1 and pin~4 are \texttt{GND} and 3.3~V \texttt{Vcc}, respectively. Pin 3 is pulled up to 3.3~V via a 4.7~K$\ohm$ resistor. We suspect it is either the reset pin or the \texttt{TCLK} signal of the cJTAG (Compact JTAG designed by MISP company) debugging interface. By trying to short pin 3 and pin 1, we observed the IP camera is reset, so we can conclude that pin 3 is the reset pin of the Realtek SoC. During the entire readout process, we need to short the reset pin, and the GND pin with tweezers to keep the SoC inoperative, as illustrated in~\autoref{fig:Camera_clip}~(b).

At this point, the binary image is dumped from the Flash chip, and the victim device can be re-assembled. No further physical access is required. Next, we extract the sensitive information\footnote{\scriptsize We follow the Blog post at \underline{https://drmnsamoliu.github.io/}}. The user configurations (password, WiFi SSID, and passphrase) are located in memory address 0x40000-0x50000 in the dumped file. However, this partition is compressed with \texttt{Zlib} and encrypted with DES (data encryption standard).

The camera must first decrypt the configurations at start-up and read the WiFi password before connecting to the Internet. Hence the DES key must be stored on-device. According to the blog post, the key is derived from a string ``C200 1.0" (corresponding to the model number and hardware version) at address 0x600c0. However, at this address, in the memory image we dumped, there is a meaningless string (0x06 0x68 0x7a 0x88 0xa8 0xa7 0x01 0x97). We know the model number of our IP camera is C100, and the hardware version is 2.0 according to the nameplate. Therefore, it is natural to seek to search for ``C100 2.0" in the dumbed file. A matched string appears at memory address 0x700c0. The sting ``C100 2.0" is the model-specific key material. To derive the correct DES key, a hash function\footnote{\scriptsize Available:~\underline{https://drmnsamoliu.github.io/assets/code/key.c}} extracted from IP camera's firmware is used. By replacing the key material ``C200 1.0" with ``C100 2.0", a 64-bit DES key ``249c6923" can be derived. Since OpenSSL takes hex strings instead of character strings, we must convert the key into hexadecimal value 0x3234396336393233. Subsequently, we can employ the following command to decrypt the dumped memory image:

\begin{lstlisting}[style=BashInputStyle]
# openssl enc -d -des-ecb -nopad -K [DES key] \
-in [dumped image] -out [out file]
\end{lstlisting}

This command specified using the encrypting function ``enc" in the OpenSSL toolbox. The parameter ``-des-ecb" specifies selecting the DES cipher and operating it in ECB (electronic codebook) mode. No padding is used as ``-nopad". The DES key 3234396336393233 should be filed after the keying flag ``-K". The dumped memory image is passed in, following the ``-in" flag. In the end, use the ``-out" flag to specify a location to save the decrypted file. Once the dumped memory image is decrypted, we can use \texttt{binwalk} command to decompress it:
\begin{lstlisting}[style=BashInputStyle]
# binwalk -e [decrypted file]
\end{lstlisting}
\texttt{binwalk} is an open-source toolbox for firmware image analysis. Flag ``-e" indicates to extract known file types automatically. The extracted user configuration will be placed in a new folder. If successful, a readable user configuration should be available.

Inside the extracted user configuration file, we can see the IP address, supported network protocols, user name, and passwords. Instead of storing a clear text password, the IP camera stores a hash value of the password. As a hash function, it is undesirable to be able to invert a hash value to its plaintext form (a.k.a., pre-image attack~\cite{sasaki2009finding}). One promising attack to revert the hash is to use a rainbow table. Rainbow table is a pre-computed mapping table from chosen plaintext to hash values and vice versa. We have used the online rainbow table website~\underline{https://crackstation.net/} to successfully revert the password of one shared account in our IP camera and used this information to successfully gain access to the RTSP (real-time streaming protocol) from the IP camera. 

We demonstrate obtaining unauthorized access to a video stream using information extracted from the WiFi IP camera's memory dump in \underline{https://youtu.be/fnIn9QugrXI}. Importantly, the entire attack process takes \textit{less than 25 minutes}. Only the \textit{first 8 minutes require physical access to the target camera}. The firmware analysis and cracking of the dumped firmware can be done offline, using freely available tools.

\vspace{1mm}
\noindent\textbf{Summary.~}The data stored in off-chip Flash memory can be easily read out through the exposed memory bus even when the system is powered off. The manufacturer has employed multiple techniques to enhance the security, such as encrypting the user configuration partition and storing hash value instead of the original password. The shared password can be reverted from hash using a rainbow table in our demo. Password salting could effectively mitigate such attacks.

\section{Conclusion and Discussion}\label{sec:conclusion}
We considered the dangers of open debug interfaces and exposed memory buses in commercial IoT devices. With two case studies, we showed the simplicity and the low cost of attacks by a person with entry-level knowledge on embedded systems--notably, the first attack only required less than 30~minutes and the second, only requires less than 8~minutes of access to the device. Evidently, security of IoT devices still require further emphasis from device manufacturers. Strategies such as disabling the debug interface supported by SoCs, securing the exposed memory buses by encrypting sensitive memory partitions and salting passwords are minimal to no additional cost step to improve current state-of-practice. However, secure on-device key derivation remains a challenging problem where memory fingerprint based methods can provide secure alternatives~\cite{maes2012pufky,su2019secucode,gao2019building}. We hope our work will help support development of mitigation methods, inform threat models, and increase awareness of a different threat dimension posed by electronic devices employed in everyday life. 

\bibliographystyle{abbrv}
\bibliography{main.bib}
\appendices
\section{Resources} \label{apd:resources}
A key resource is a book by reverse engineering expert Edwin Sobey\footnote{E. Sobey. \textit{Unscrewed: Salvage and Reuse Motors, Gears, Switches, and More from Your Old Electronics}. Chicago Review Press, 2011.}. Sobey talks about salvaging useful components from broken or old electronics in this book. More importantly, readers can study how to use correct tools to dissect the device, identify valuable parts that can be reused, basic knowledge about things they work, and safety rules to follow when unscrewing devices. Notably, this book was published in 2011 and some techniques are outdated with the rapid evolution of industrial designs. For example, nowadays, more and more devices have their casing held together with adhesive or ultrasound welding rather than screws and clips commonly used in 2011. For readers without an electronics background, an introductory book to build and test some simple circuits in \textit{Electronics for Beginners} written by Jonathan Bartlett~\cite{bartlett2020electronics} is a good starting point. This book covers basic electronics concepts, schematics, circuit analysis and calculations. Given some familiarity with electronics, \textit{The IoT Hacker's Handbook} written by Aditya Gupta~\cite{gupta2019iot} provides in-depth insights into hardware and embedded system exploitation to firmware exploitation. 

Then, iFixit\footnote{www.ifixit.com} is a wiki-based website for user-generated content sharing focusing on repairing technological devices. The website provides video content from experienced people with step-by-step recipes to tear down the device. This may largely reduce the risk of having your device damaged or injuring yourself. Another wiki-based website, exploitee.rs, focuses on hacking where the content provides information on how to find hidden debug interfaces and how to decrypt dumped firmware from the examples provided therein.

\section{Attacker tools}\label{apd:attacker_tools}
\begin{itemize}
    \item \textbf{Multimeter:} An instrument that can measure basic electrical properties, such as voltage, current, resistance, and so on. We recommend choosing a multimeter with diode forward biasing and wire connectivity function. The model we used in this work is a US\$33 Stanley STHT77364, an US\$10 alternative Gator XL830L from could be an alternative.
    \item \textbf{Logic analyzer:} An instrument that can measure fast varying digital signals, records those signals over time domain, and performs analysis to discover the information encoded. In this work, we did not use a logic analyzer, in future work we will use a US\$300 Digilent Analog Discover 2 to demonstrate the analysis of U-Boot entry point by monitoring the off-chip Flash memory bus traffic. For such kind of task a 24MHZ 8 Channel open-source logic analyzer priced US\$15 is adequate.
    \item \textbf{Flash memory programmer:} A low-cost (US\$15) device to read out the Flash (generally also supports EEPROM) memory content from or write image files to Flash memory chips. A model with a test clip for fast and clean hooking up the exposed memory bus is more useful. The CH341A Pro Flash memory programmer used in this work is priced US\$10 on ebay.com.
    \item \textbf{JTAG programmer/debugger:} JTAG is also known as the IEEE 1149.1 standard for deploying and debugging firmware on the chip and also offers low-cost and time-saving testing for all components in a system through boundary-scan. JTAG is widely used in industry. Different system architectures may require different JTAG programmers, the one we used to extract electronic lock programming code from MSP430G2433 is an US\$150 MSP-FET430UIF. Compatible MSP430 JTAG programmer from a third party is around US\$35. An universal JTAG programmer supports ARM, MIPS and RISC-V also priced at US\$35 on ebay.com. 
    \item \textbf{Embedded system development board:} Such as Raspberry Pi Zero (US\$10), Arduino UNO R3 (US\$22) or STM32 Bluebell (US\$15). Those development boards are useful, for example, to deploy JTAGenum\footnote{\underline{https://github.com/cyphunk/JTAGenum}}, an open-source program for identifying JTAG pin-out definitions.
\end{itemize}

\section{Disassembly of The Electronic Lock and IP Camera}
\label{apd:Disa_Elec_lock}\label{apd:Disa_IP_cam}
\begin{figure}[h]
    \centering
    \includegraphics[width=\linewidth]{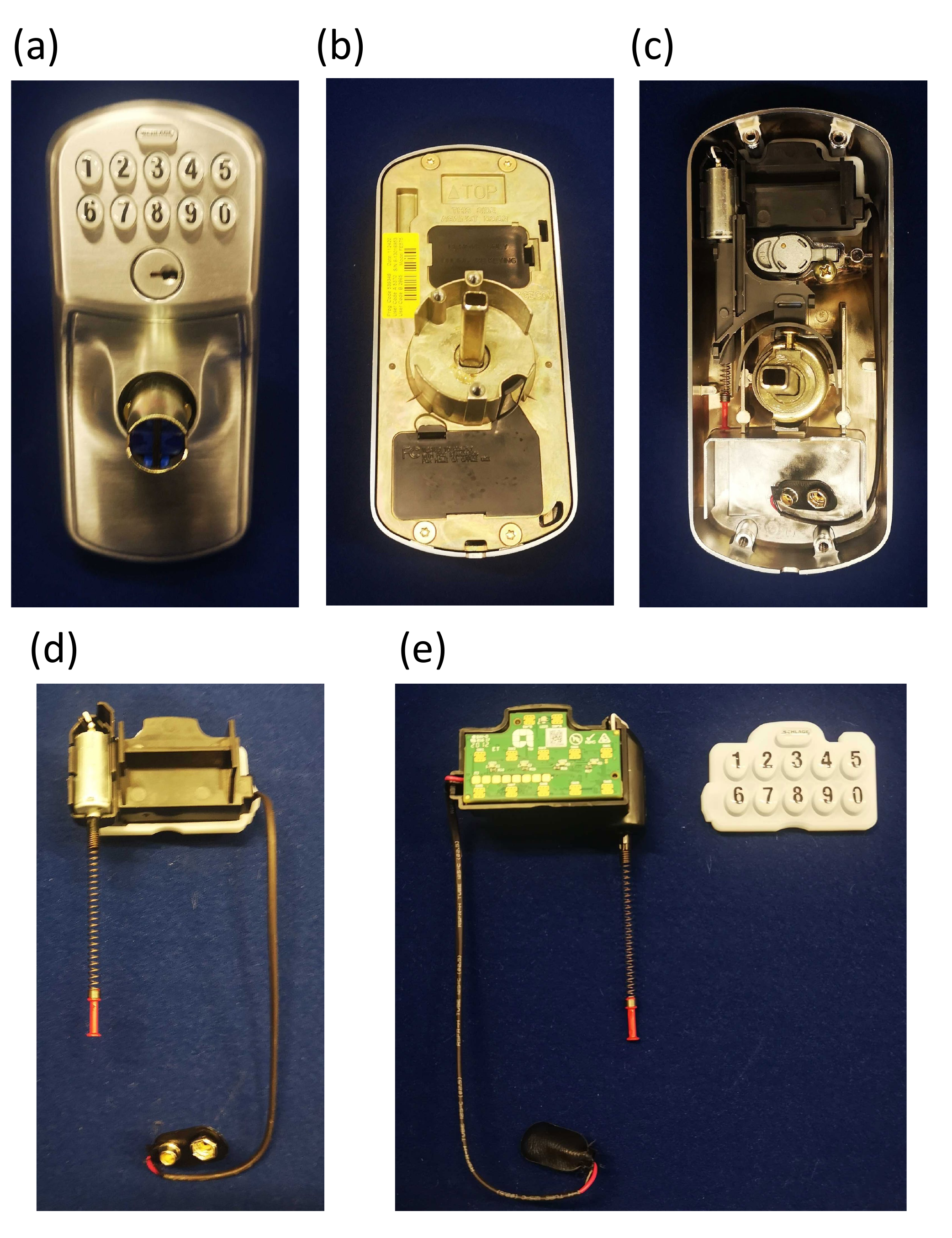}
     \caption{Tear-down of the electronic lock: (a) front side; (b) back side; (c) with back panel removed; (d) the electronic assembly; (e) the circuit board is visible after removing the waterproof silicone rubber keyboard.}
    \label{fig:Lock_teardown}
\end{figure}

To disassemble the electronic lock, a T-10 screwdriver is required to remove the four bolts holding the back panel in place. Subsequently, all internal parts can be removed by hand within 1~minute, without a tool, as illustrated in~\autoref{fig:Lock_teardown}. Our technique to get access to the JTAG pins is to solder a 1.27~mm 7-pin header. This requires 5~minutes. Accessing the internal NVM using MSP-FET430UIF debugger can take another 3~minutes.  We can de-solder the jumper wire header in 3~minutes with a hot air gun and take another 3~minutes to put all parts together. The entire process takes 15~minutes and could be further shortened by using a data repair tool (available on \textit{ebay.com} for US\$21) instead of soldering a pin header.

\begin{figure}[h]
    \centering
    \includegraphics[width=.7\linewidth]{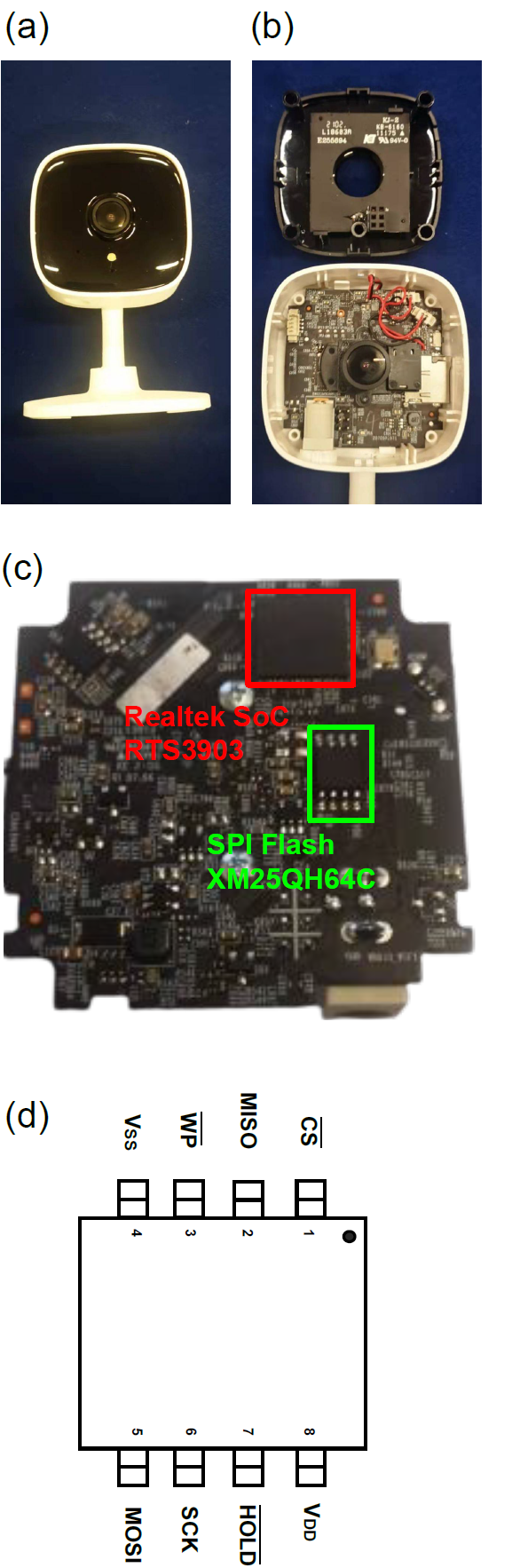}
     \caption{(a) the front side of the TP-link Tapo C100 IP camera; (b) with the front panel removed; (c) the back side of the main board; and (d) the pin definition of the MX25QH64 SPI Flash chip.}
    \label{fig:TapoC100}
\end{figure}

The disassembly of the WiFi IP camera simpler. We can remove the front panel of the camera by inserting a lever into its edge and gently applying a force as shown in \autoref{fig:TapoC100}. Then the mainboard is removed by freeing the snap-fit in less than one minute. It takes tens of seconds to attach the Flash programmer test clip to the exposed Flash chip and short circuit the \texttt{RESET} pin of the SoC. Reading the Flash memory via SPI is the most time-consuming part, taking 3 to 4~minutes. Once the memory image is dumped, the target IP camera is re-assembled in another minute. We conclude the entire process is possible in 8~minutes. The firmware analysis and cracking of the dumped firmware can be done offline, as shown in our demo video \underline{https://youtu.be/fnIn9QugrXI}.

\end{document}